# Cladding Layer Enhanced GHz Bulk Acoustic Wave Resonance in Sodium Niobate Thin Films on Silicon

*Zhi Shiuh Lim*, Qibin Zeng, Hui Kim Hui, Mengyao Xiao, Tiancheng Luo, Weifan Cai, Shengwei Zeng, Samantha Faye Duran Solco, Baichen Lin, Celine Sim, Zhen Ye, Jinlong Xu, Mingxi Chen, Wei Fu, Chee Kiang Ivan Tan, Seeram Ramakrishna, Yeng Ming Lam, Vincent Chengkuo Lee, Ariando Ariando, Huajun Liu**

Zhi Shiuh Lim, Qibin Zeng, Hui Kim Hui, Tiancheng Luo, Weifan Cai, Shengwei Zeng, Samantha Faye Duran Solco, Mingxi Chen, Wei Fu, Chee Kiang Ivan Tan, Huajun Liu

Institute of Materials Research and Engineering (IMRE), A*STAR, 2 Fusionopolis Way, Innovis, #08-03, Singapore 138634.

Mengyao Xiao, Jinlong Xu, Vincent Chengkuo Lee

Department of Electrical & Computer Engineering, National University of Singapore, Block E4 Level 5, Room 42, 4 Engineering Drive 3, Singapore 117583.

Baichen Lin, Celine Sim, Yeng Ming Lam, Huajun Liu

School of Materials Science and Engineering, Nanyang Technological University, Blk N4.1, Nanyang Avenue Singapore 639798.

Zhen Ye, Seeram Ramakrishna

NUS Mechanical Engineering, College of Design and Engineering, National University of Singapore, 9 Engineering Drive 1, Block EA #07-08, Singapore 117575.

Ariando Ariando

Department of Physics, Faculty of Science, National University of Singapore, Block S12 Level 2, 2 Science Drive 3, Singapore 117551.

Email: lim_zhi_shiuh@a-star.edu.sg, liu_huajun@a-star.edu.sg

Funding: Competitive Research Programme (CRP), Singapore National Research Foundation (NRF)

**Keywords:**

Sodium niobate, piezoelectric acoustic resonator, polarization anisotropy, countering thermal tensile strain

**Abstract:**

Bulk acoustic wave resonators (BAWR) and bandpass filters operating at GHz frequency are the workhorse of (Vo-)LTE telecommunication and broadband internet. In line with the Singapore Green Plan 2030 for innovating environmentally friendly products, we fabricated lead-free BAWR with sodium niobate ($NaNbO_3$) piezoelectric on silicon with a high electromechanical coupling factor up to 31.3% operating at ~4 GHz. We disclose our cladding-layer strategy in mitigating leakage current and avoiding cracks induced by thermal strain, by cladding $NaNbO_3$ in between high band gap perovskite oxide insulators. In addition to the objectives mentioned, we also verified the efficacy of reducing lattice parameters of the cladding layers in promoting vertically distorted tetragonal phase $NaNbO_3$ and producing stronger BAWR signals.

## 1. Introduction

The crystal structure of perovskite oxides ($ABO_3$) is famous for hosting varieties of piezo- and ferroelectric compounds. The simple atomic structure of perovskites also offers large tunability ferroic orders by strain engineering. One environmentally friendly piezoelectric material – $NaNbO_3$ (NNO) – remains less well-understood due to the existence of multiple non-centrosymmetric structural phases. Several prominent phases, namely, the ferroelectric (FE) $Pmc2_1$ in (001), ferroelectric $R3c$ in (111) and antiferroelectric (AFE) *Pbcm* are known to have very near (<5 meV/formula unit) formation energies. In recent years, such multi-phase coexistence has led to the discovery of giant effective piezoelectric coefficient ($d_{eff,33}$) up to astounding magnitudes of several nm/V[1] contributed by self-assembled nanopillars and structural heterogeneity. It also encourages the idea of voltage-induced phase transition leading to large tuning effects, which are attractive properties for fabricating radiofrequency microelectromechanical systems (RF-MEMS). Specifically, in line with the surge of development in artificial intelligence, NNO has the potential for developing a next-generation tunable acoustic filter for the 5G/6G broadband internet to cope with the increasing demand for large bandwidth telecommunications. To qualify for large-scale industrial production, it is necessary to build NNO-based RF-MEMS on silicon wafers, since silicon is the most widely used, cost-effective, contamination free and low acoustic-loss platform.

Pure $NaNbO_3$ with perfect stoichiometry and oxidation states is known to have an adequately high indirect electronic bandgap of 3.2-3.4 eV[2]. However, a physical vapour deposition environment has a high tendency for sodium deficiency accompanied by a difficulty in reaching full ionization of $Nb^{5+}$[3]. The ubiquitous mixture of $Nb^{4+}/Nb^{5+}$ causes unintentional hole-doping and effective bandgap reduction, leading to the small polaron hopping leakage current mechanism[4]. On the other hand, for acoustic resonators, it is vital to simultaneously fulfill both the requirements of high electrode conductivity (~$10^7$ S/m) and low leakage current in the insulator layer (≤1 nA). This is evident from the success of the commercial BAWRs made of AlN and Sc-doped AlN sandwiched between Mo electrodes achieving quality (Q) factor up to 10000, even though its effective piezoelectric coefficient ($d_{33}$≈3-5 pm/V) is not high[5]. In this work, we show that the mentioned requirements can be fulfilled in NNO by using lattice-matched epitaxial cladding layers made of wide bandgap (~5.6 eV) insulators (WBI) to compensate for its large leakage current, leading to a significant enhancement of piezoelectric resonance signals. The cladding layers also suppress the formation of cracks that have been ubiquitous among NNO films grown on silicon due to the large mismatch in thermal expansion coefficients[6]. We also show that by the choice of a perovskite cladding layer with smaller lattice parameters, it is possible to promote the vertically aligned tetragonal phase of NNO despite that the cladding layers (~20 nm) are much thinner than the NNO film thickness (~250 nm). This is evidenced by a larger c-axis lattice parameter in $NaNbO_3$, which is consistent to a stronger BAWR signal observed that primarily relies on $d_{eff,33}$.

**Thin film characterizations**

Firstly, we explored the $ZrO_2$-$CeO_2$-$La_{0.5}Sr_{0.5}CoO_3$ tri-layer buffer method[7] (henceforth abbreviated as the ZCL-route) for growing perovskite on Si(001), as shown in Fig. 1a. To bridge the lattice mismatch, the crystal lattice of perovskite layers, starting from $La_{0.5}Sr_{0.5}CoO_3$ (LSCO), would naturally form a 45° in-plane rotation with respect to the larger cubic lattice below. Besides, in a physical vapour deposition (PVD) system without reaching ultrahigh vacuum (UHV), the $ZrO_2$ film has gained fame as a great candidate for scavenging oxygen from the surface native oxide ($SiO_x$). Thus far, several balanced attributes are believed to contribute to the easier success of epitaxial $ZrO_2$//Si(001), namely, the low enough electronegativity of Zr (1.33) on the Pauling scale, and the tendency to form oxygen vacancy on the target pellet surface at high vacuum and after laser/plasma ablation which potentially maximizes its “getter” ability. In comparison, albeit the Sr and SrO buffers have an even lower electronegativity, UHV is still mandatory for their successful two-dimensional epitaxial growth[8]. Further details are described in *methods*.

As shown in Figure 1b (bottom panel), the deposition was started with a Si(001) surface untreated with buffered hydrofluoric acid (BHF), hence the native oxide ($SiO_x$) is present. We found that using BHF resulted in a brighter/sharper RHEED pattern of the Si(001) but would promote the formation of near-amorphous $ZrO_2$(111) and $CeO_2$(111) in our PLD environment and hence is undesirable (Supporting Figure S1a). A sufficient waiting time or slow deposition rate of $ZrO_2$ is needed for the reaction between $ZrO_x$ and $SiO_x$, therefore producing a blurry but streaky RHEED pattern. After sufficient thickness, $CeO_2$(001) can grow well on $ZrO_2$(001) with a bright-spotty RHEED pattern indicating high-quality island growth mode, while maintaining the r.m.s. roughness at ~0.5nm. Next, the LSCO has been found by Tanaka's group[7, 9] to be the excellent perovskite that facilitates the mentioned crossover of crystal structure achieving a rocking curve full-width-half-maximum ($\Delta\omega_{FWHM}$) of 1.1° in X-ray diffraction (XRD). Subsequent film growths of thick $SrRuO_3$ (SRO) and NNO films as the bottom electrode and piezoelectric layer are straightforward, showing a progressively reducing $\Delta\omega_{FWHM}$ down to 0.34° (Fig. 1c inset) and a transformation of RHEED pattern back to bright-streaky. $SrRuO_3$ was chosen as the bottom electrode since it is the most conductive perovskite oxide known to-date. We also provide clear evidence of the 45° rotation by pole figure measurements (in $\chi$, $\phi$ domains) of the NNO(013) and Si(026) family of planes as shown in Figure 1d, e.

An improvised buffer layer system coined as the "ZPS-route"[10] was also explored, by sputtering crystalline Pt(001) directly on $ZrO_2$ at an elevated temperature as shown in Fig. 2a. Integrating Pt(001) into the ZCL-route has yet to be successful at present (Supporting Figure S1a). On the other hand, the face-centred cubic (FCC) lattice of Pt(001) does not rotate by 45° with respect to that of $ZrO_2$ despite high lattice mismatch, hence resulting in the pole figure of NNO(013) and Si(026) as shown Fig. 2d, e. Analysis of the perovskite $2\theta$-$\omega$ XRD must be done around (001) or (003) since the strong (002) peak contributed by the 150 nm-thick Pt film would obscure the perovskite films' peak around (002), but Pt(001) does not exist due to destructive interference by the FCC structural factor. The final films of Pt, $SrRuO_3$ and NNO show $\Delta\omega_{FWHM}$ reduction in sequence of 0.7°, 0.6° and 0.35°, achieving similar quality as the ZCL-route mentioned. Although the $SrRuO_3$ is not needed as the bottom electrode, it functions as a buffer for quality improvement as well as a barrier/sacrificial layer preventing Pt diffusion into the insulating perovskite $NaNbO_3$, which has been a well-known problem. This route offers the advantage of reducing the bottom electrode's electrical resistivity by around 2 orders of magnitude.

We further explored the effect of inserting two WBI cladding layers that sandwich the NNO, namely, $DyScO_3$ (DSO) or $LaAlO_3$ (LAO) (Supporting Figures S3 and S4), with their thicknesses kept thin (~20nm) to avoid competing excessive voltage drop with NNO. In addition to having wide bandgaps, DSO and LAO were deliberately chosen for their complete absence of piezoelectricity and largely different pseudocubic lattice parameters (3.95 Å and 3.79 Å respectively) to explore their strain effects on NNO. Fig. 2c shows that the NNO(001) peaks are obviously split comparing the two sandwiches DSO/NNO/DSO (DND) and LAO/NNO/LAO (LNL). Surprisingly, the very thin cladding layers can exert a significant in-plane strain on the thick NNO film. The NNO's c-axis lattice parameters ($c_{NNO}$) were found to be 3.89 Å and 3.93 Å for DND and LNL respectively, suggesting that the LNL sandwich is more suitable for FBAR fabrication due to its vertical tetragonal domains. Likewise, the NNO film also exerts in-plane strain onto the two cladding layers, resulting in $c_{DSO}$ = 4.04 Å (elongation) and $c_{LAO}$ ~3.77 Å (contraction). The mentioned concept is consistent to the selected area electron diffraction (SAED) results done in a transmission electron microscope (TEM) as shown in Supporting Figure S1c-d.

**Device fabrication and results**

Subsequently, several heterostructures were tested by a vector network analyzer (VNA) for resonance signals via an RF probe in a single-port configuration, after completing the photolithography, top Pt electrode deposition and lift-off steps. Our lithography design of circular (50 μm diameter) S-pads disconnected from the surrounding large area G-pad ensures proper grounding. An optional DC bias would be added on top of the small (0.5 $V_{pp}$) AC measurement voltage to polarize the NNO film as shown in the Fig. 4a measurement schematic (top-left corner). Firstly, as shown in Fig. 3b, the DC bias dependent VNA data obtained from the ZPS-route without WBI-claddings for NNO was found to be very weak, i.e. maximum -0.66 dB in $S_{11}$ and 1.2 dB-Ω in $Z_{11}$ at frequency ~3 GHz and DC bias of -10 V. We identified that NNO films grown on Si(001) is prone to developing crack (Figure 3a) due to large mismatch in thermal expansion coefficient, i.e. up to 1.6x$10^{-5}$ in NNO but 2.6x$10^{-6}$ in silicon. The crack is the primary reason for the poor resonance due to non-uniform vibration, high leakage current and a low DC voltage limit before breakdown. On the other hand, the ZCL-route without Pt resulted in a maximum resonance signal of $S_{11}$ ~-1.2 dB and $Z_{11}$ ~2.7 dB-Ω as shown in Fig. 3c, but the undesirable broad peaks are well-known resulted from a high electrical loss (low Q-factor) due to poor electrode conductivity. With the presence of WBI-claddings in DND- and LNL-sandwiches, the crack issue has been avoided, and the resonance signals are largely improved with narrower peaks as seen from Fig. 3d,e. The LNL-sandwich recorded a resonance signal of $S_{11}$~-5.9 dB and $Z_{11}$~8.4 dB-Ω at 3.4 GHz and DC bias of -30 V, significantly stronger than that of the DND-sandwich and is currently our best result. Similar values of electromechanical coupling factor $k_{eff}^2 = \frac{\pi^2}{4}\left(\frac{f_a - f_r}{f_a}\right)$ of 31.3% and 29.8% were found for the DND and LNL structures respectively. However, the quality factor (defined as $Q = \frac{f_{\mathrm{res}}}{\Delta f_{\mathrm{FWHM}}}$) was low, i.e. 10.8 and 9.6 for the DSD and LNL structures respectively.

We further compared the remanence signals from the DND- and LNL-sandwiches by measuring the resonance after removing the high DC voltages. Fig. 4c shows that the LNL shows a remanence signal of $S_{11}$~-0.7 dB and $Z_{11}$~1.3 dB-Ω after removing the negative DC bias -30 V (negative remanence), but is negligible after removing the positive DC bias of +20 V. The magnitude of negative remanence signal also coincides with the signal while the DC +20 V is applied, suggesting a wake-up effect to be investigated in the future work. However, any remanence signal is completely absent in the DND, as shown in Fig. 4b. The laser Doppler vibrometer (LDV) setup as shown in Fig. 4a (bottom-right) is an indispensable tool for measuring the sample's effective piezoelectric coefficient ($d_{33,eff}$) at an amazing picometer resolution at 1 MHz and room temperature. The $d_{33,eff}$ of the DND- and LNL-sandwiches were found to be ~6 pm/V and ~20 pm/V respectively. Hysteretic butterfly loops are also noticeable in the LNL but are absent in the DND, indicating that the DND is lack of an out-of-plane ferroelectric polarization. The $d_{33,eff}$ of LNL is only slightly lower than ~30 pm/V measured from NNO films grown on single crystal 0.1%Nb-doped $SrTiO_3$ (NbSTO) (001) substrates which applies an in-plane compressive strain onto NNO.

**Discussions**

In an overview, using NNO in FBAR has demonstrated significant challenges. Firstly, although the ZPS-route offers a real metallic Pt bottom electrode, the lack of 45° degree in-plane rotation of FCC lattice further enhances the tensile strain on perovskites, hence leading to a higher tendency of cracks in NNO compared to the ZCL-route. At present, our heterostructures qualify as a solidly mounted resonator (SMR-BAW) without well-matched Bragg reflectors; a true FBAR with acoustic cavity below the device has not yet been formed by releasing the silicon substrate. This is evident from our resonance signals containing a low fraction of high-overtone bulk acoustic resonance (HBAR) signal where multiple modes of standing waves are formed within the Si(001) substrates. It is well known that successful creation of an acoustic cavity in FBAR would result in a large improvement in the Q-factor. However, during silicon substrate release by $XeF_2$ gas etching, the fragile NNO films present a higher tendency of collapse compared to other piezoelectric material and is left as a future work that requires more meticulous attention. Likewise, the accidental Bragg reflectors seen in Fig. 3 formed by the ZCL- and ZPS buffer layers do not have well-matched thickness to the quarter acoustic wavelength ($\lambda_a/4$). Our efforts here primarily aim to improve the NNO's performance and has not reached a condition for acoustic filter fabrication.

From historical effort, it has been known that the in-plane structural and thermal strains acting on perovskite films by the Si(001) substrate are usually in competition. For example, the structural strain is usually compressive by comparing a ferroelectric perovskite with lattice parameter (>3.9 Å) to the 45°-rotated effective lattice parameter of Si(001) of $5.431/\sqrt{2}$=3.84 Å, and it is dominant only at an ultrathin (first few atomic layer) of perovskite away from the interface. In contrast, the thermal strain occurs during sample cool down in the growth chamber, it is always in-plane tensile and would dominate at large film thicknesses. This has been clearly confirmed in Ref. [8c, 11] where thick $SrTiO_3$ films showed smaller-than-bulk c-axis lattice parameters. In our case with the ZCL- and ZPS-routes (unlike the direct growth by MBE), since the thick NNO films are not in proximity with the oxide/Si interface, the structural compression may be completely absent, hence the thermal tensile strain would always favour an in-plane polarization anisotropy. Nevertheless, this work shows that a wise choice of WBI cladding layers with small lattice parameter is a good strategy to counter the thermal strain and restore the vertical remanence polarization to some extent. Newer cladded structures could be designed in the future to further lower the crystal symmetry, such as enhancing the super-tetragonality (*P4mm*) of a perovskite piezoelectric film in general, without compromising the crystal quality or excessively increasing the mass loading. After the successful creation of acoustic cavities below FBAR devices by etching the silicon wafer, the $Z_{11}$ signal and Q-factor may enhance up to >20 dB and ~100 respectively, approaching the industry standard for making the next generation acoustic filter for broadband internet.

**Methods**

The Si(001) wafer was first cleaned by ultrasonication in acetone without any hydrofluoric (HF) acid treatment, since removal of the ~2 nm-thick native oxide layer ($SiO_x$) on the wafer surface was proven to promote the formation of poor quality $ZrO_2$(111) and $CeO_2$(111). Oxide film growth was performed in a PLD chamber and a Coherent COMPexPro 110F KrF excimer laser. The chamber is equipped with an RHEED system consisting of a K-space Associate KSA400 camera and a STAIB electron gun operating at 25 kV and 1.5 Amps of filament current. The Si(001) wafer was then heated to 750 ºC in the PLD chamber's base vacuum of $1x10^{-6}$ Torr. Hence the native $SiO_x$ layer is not expected to thermally desorb during the heating process, instead, it can be understood to thicken slightly since the (2x1) reconstruction RHEED pattern of Si(001) surface will become dimmer during the heating process. Then, the crucial "native oxide scavenging" step was begun by supplying $ZrO_2$ plume up to a thickness of 2 nm at the base vacuum and 750 ºC, resulting in complete blurring of the RHEED pattern. After a sufficient waiting time of 10 minutes, a streaky RHEED pattern would form. Oxygen gas was then introduced into the chamber to reach a pressure of $5x10^{-4}$ Torr at a slow rate to prevent overshooting, and the $ZrO_2$ deposition continued to reach a total thickness of 46 nm while maintaining the streaky RHEED pattern. The ZCL-route would continue with the $CeO_2$ deposition was done at 750 ºC and 2 mTorr $O_2$ up to 26 nm thickness, where the RHEED pattern can be observed to change into a bright spotty array, indicating high quality island growth but with low surface roughness. The ZCL-buffer was completed by 33 nm LSCO grown at 620ºC and 15 mTorr $O_2$. On the other hand, the ZPS-route would continue with *in-situ* sample transfer into a DC sputtering chamber for the Pt deposition in an argon pressure of $1.8x10^{-3}$ mbar and temperature of 650 ºC. Both the LSCO and Pt surfaces from the two routes are suitable for the subsequent growths of SRO at 650 ºC, 50 mTorr and NNO at 700 ºC, 150 mTorr $O_2$. The NNO plume was generated by laser ablation on a "sodium-deficient" $Na_{0.93}NbO_{3-x}$ target. The cladding layers DSO and LAO were deposited by laser ablation of single crystalline substrates as the targets and with the common parameters of 700 °C and 0.5mTorr $O_2$. All completed heterostructures were annealed in the PLD chamber at 600°C for 10 hours and 250 Torr of pure oxygen gas to remove oxygen vacancies. Topography scans and RHEED patterns were checked after the completion of each layer. Another group of RHEED patterns obtained from aligning the electron beam parallel to the wafer's corner [100] are shown in Supporting Figure S1b. The parameter optimizations were done carefully to ensure that all surfaces were maintained at ≤1 nm roughness, as shown in Supporting Figure S2.

The topography scans were done by Bruker Dimension Icon Atomic Force Microscopy (AFM), and X-ray diffraction (XRD) by Rigaku SmartLab Diffractometer. Standard photolithography was done with a Karl SUSS MJB4 UV mask aligner. The RF resonance signals were collected by a Rohde and Schwarz Vector Network Analyzer (VNA), represented in terms of a single-port impedance ($Z_{11}$) and scattering ($S_{11}$) parameter for energy reflectance with frequency range 0.1-20 GHz. The signal from the LDV controller was analyzed by a Zurich Instruments MFIA lock-in amplifier at frequency 1 MHz. Both the VNA and LDV were connected to a Keithley 2450 via a bias-tee for simultaneous application of a DC voltage and AC measurements as shown in Figure 4a. TEM sample preparation was by a Helios 600 Focused Ion Beam (FIB). A Titan TEM equipped with a 200 keV electron gun and a Talos TEM equipped with an energy dispersive X-ray spectroscopy (Super-X EDS, 4 SSD) were used to obtain the cross-sectional imaging as well as label each material layer in the DND and LNL heterostructures, as shown in Supporting Figure S3 and S4.

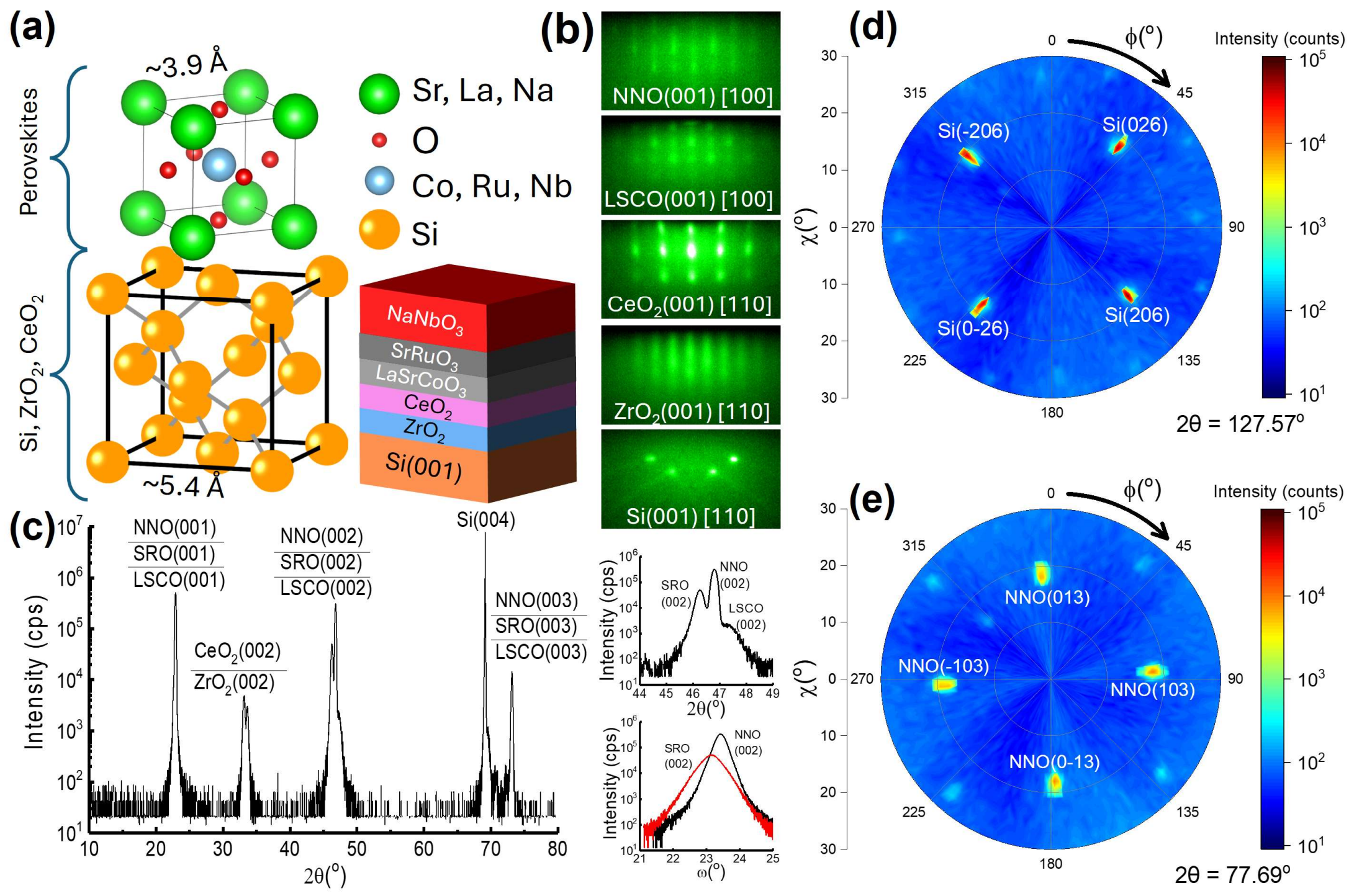

**Figure 1: Single crystalline film growth on Si(001) via the ZCL-route. (a)** Schematics of the atomic structure involving a 45° in-plane rotation and the resonator heterostructure made via the ZCL-route. **(b)** The RHEED pattern evolution of the sample surface after the completion of each layer, from bottom to top, with the wafer edge ([110]-direction) aligned to the electron beam. **(c)** 2θ-ω XRD data of the heterostructure with the ZCL-buffer, with zoomed-in (002) peaks and rocking curves shown in the right panels. **(d)-(e)** Pole figures of Si(026) and NNO(013) family of planes showing the atomic lattice alignment with a relative 45° ϕ-rotation.

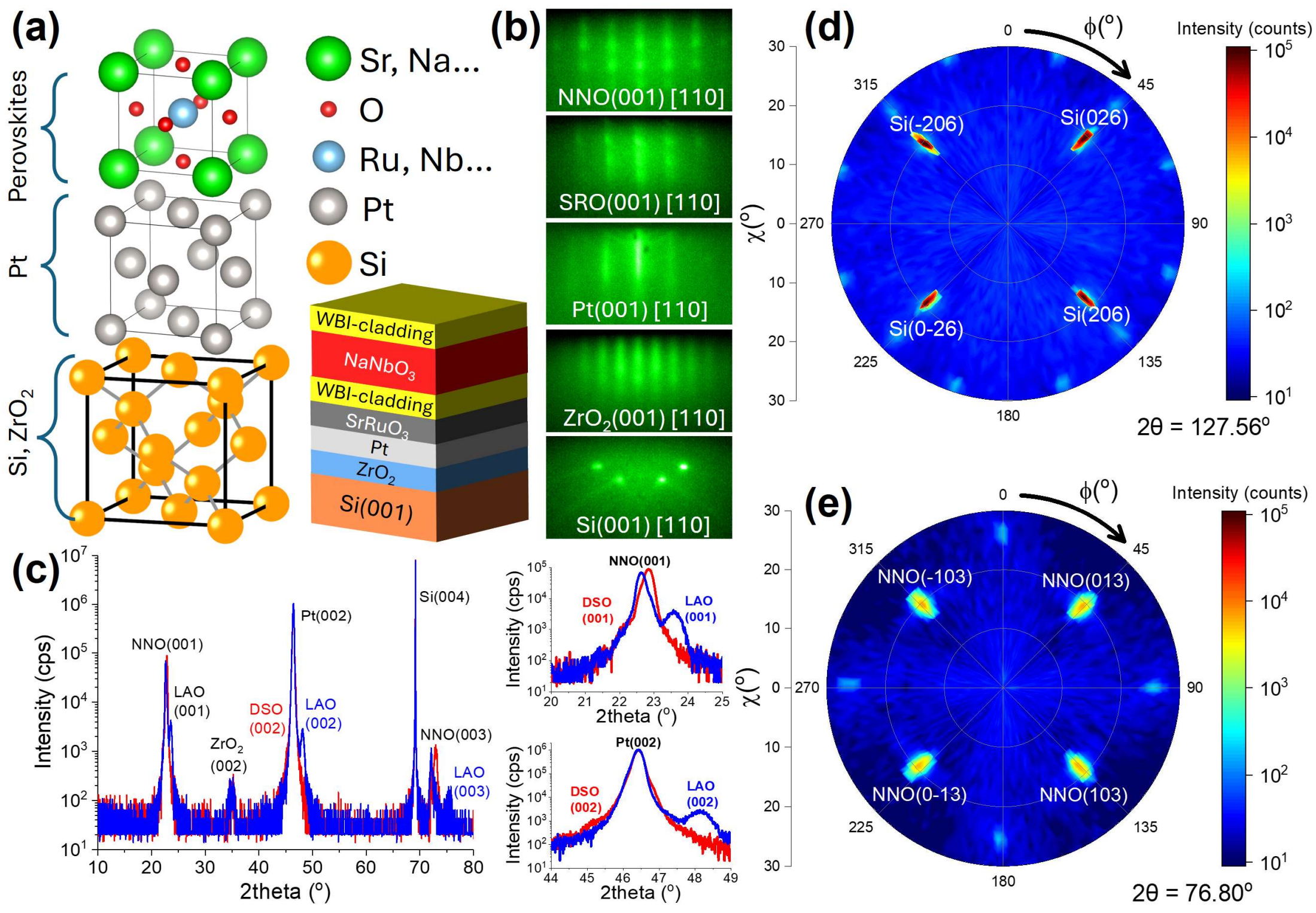

**Figure 2: Single crystalline film growth on Si(001) via the ZPS-route. (a)** Schematics of the atomic structure without in-plane rotation and the resonator heterostructures made via the ZPS-route. **(b)** The RHEED pattern evolution of the sample surface after the completion of each layer, from bottom to top, with the wafer edge ([110]-direction) aligned to the electron beam. **(c)** 2θ-ω XRD data of the heterostructure with the ZPS-buffer, with zoomed-in (001) and (002) peaks shown in the right panels. **(d)-(e)** Pole figures of Si(026) and NNO(013) family of planes showing the atomic lattice alignment without a relative ϕ-rotation.

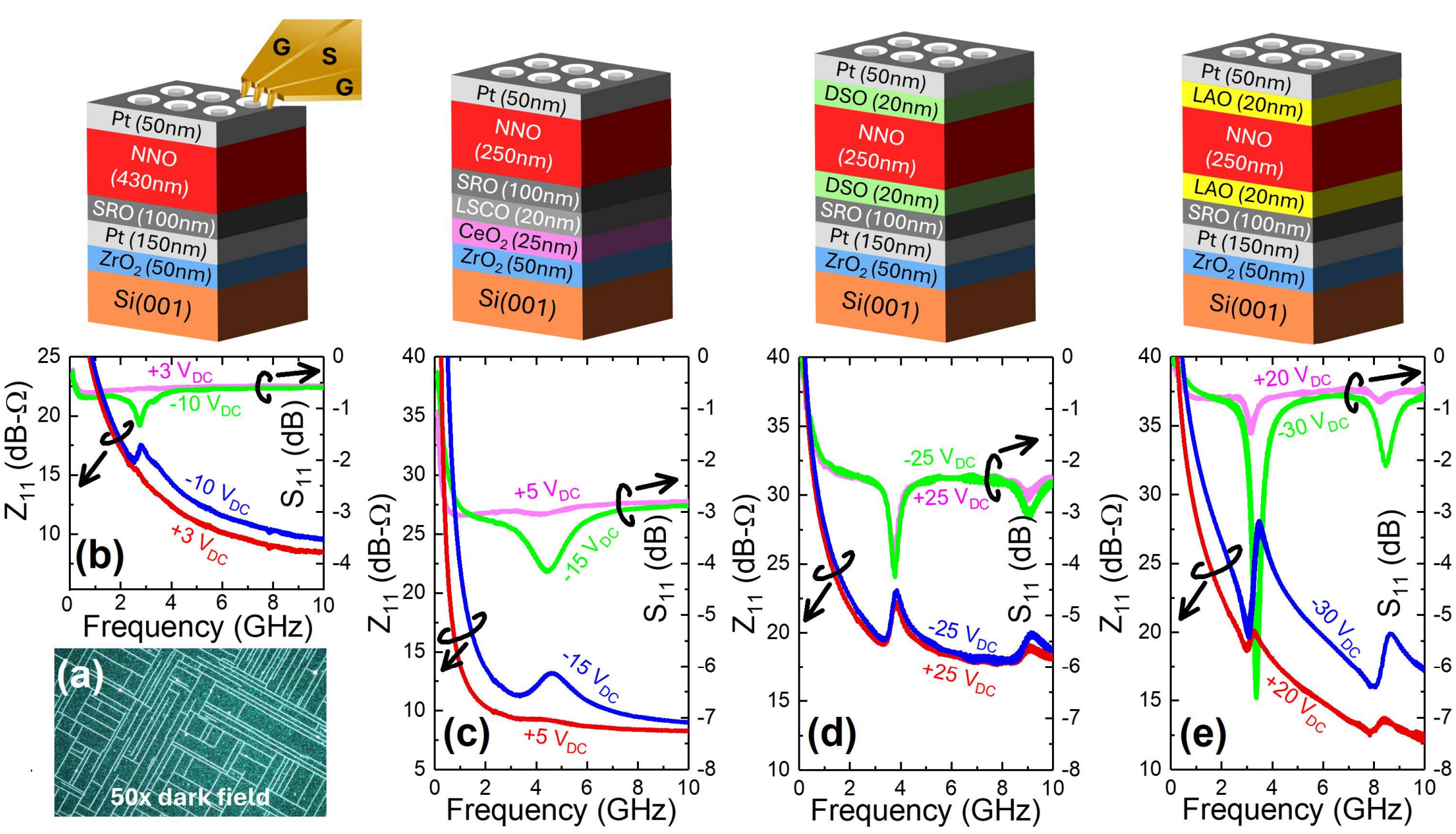

**Figure 3: GHz-BAWR signals of the NNO-based heterostructures in chronological order of development. (b)-(e)** BAWR signals represented by single-port scattering ($S_{11}$) parameter and impedance ($Z_{11}$), measured from the corresponding heterostructures shown in the respective top panels, with the top Pt electrodes patterned into circular GSG devices that matches with the size of a typical RF probe. The DC poling voltages labelled show the maximum values applicable to the device simultaneously during AC measurements without breakdown. **(a)** 50x dark-field optical image of the NNO-heterostructure built via the ZPS-route without cladding layers, corresponding to the resonance data in **(b).** White lines are cracks similar to those shown in Supporting Figure 2a. **(c)** The ZCL-route without a Pt(001) bottom electrode resulted in a broad (low-Q) resonance. **(d)-(e)** the DND and LNL sandwiches with wide bandgap claddings made via the ZPS-route resulted in higher-Q resonance and acceptable magnitudes.

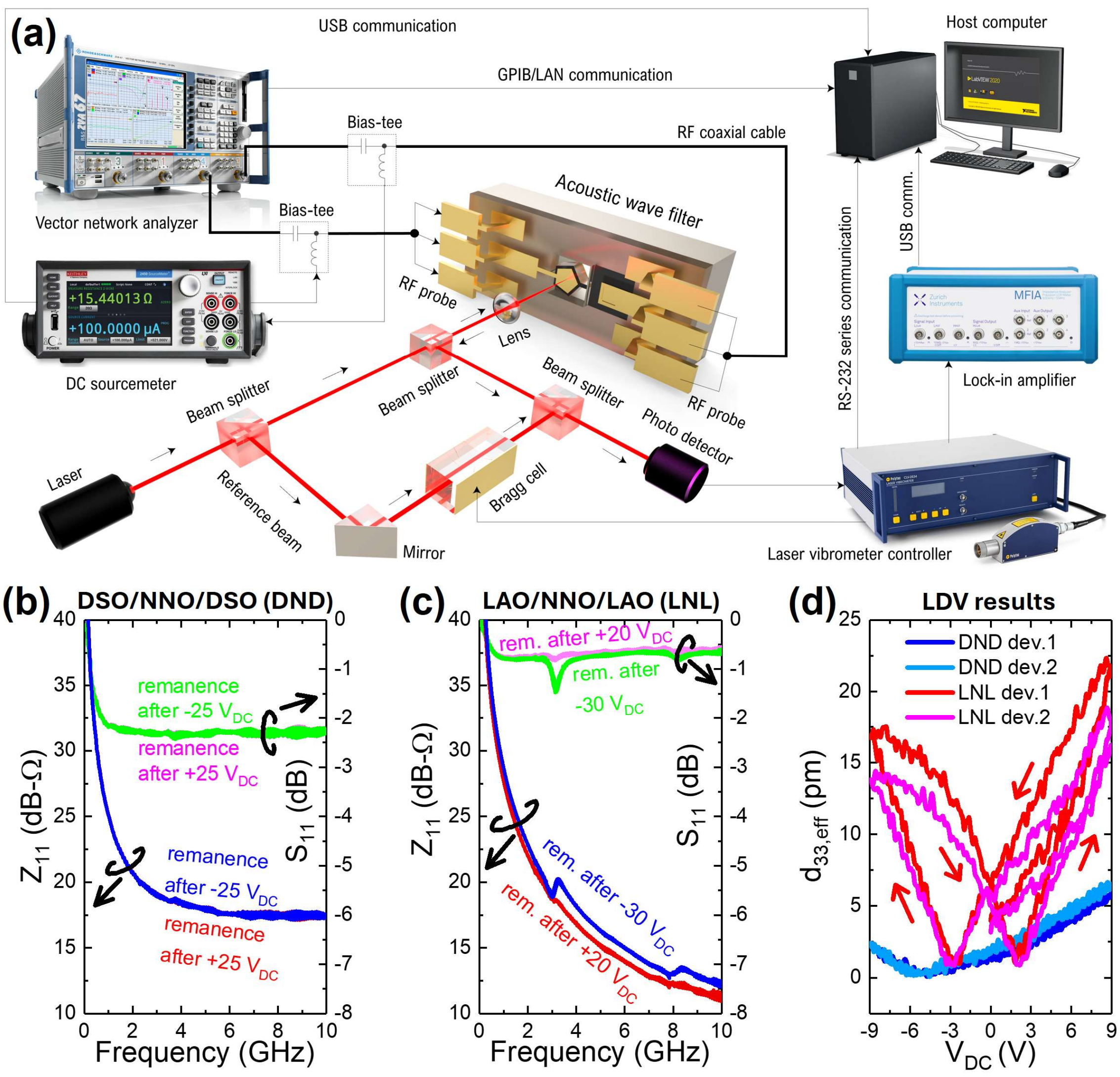

**Figure 4: Remanence resonance signals and strain-voltage loops. (a)** Schematic of the measurement circuits and computerized data collection, inclusive of a GHz RF branch (top left) and a MHz vertical mechanical vibration branch (bottom right). **(b)-(c)** the remanence BAWR signals obtained from DND and LNL heterostructures at zero $V_{DC}$ after returning from a high $V_{DC}$ history. **(d)** Mechanical vibration amplitude versus $V_{DC}$ loops for two pairs of DND and LNL devices chosen at random, where the AC measurement input was kept at 2 $V_{pp}$ at 1 MHz.

**Acknowledgements**

Z. S. L. and W. C. conceived the idea. Z. S. L. and H. L. drafted the manuscript. A. A. and H. L. provided the experimental equipment. Z. S. L. and S. Z. performed the heterostructure film growth, data collection of AFM and RHEED, and photolithography. Q. Z. constructed the electrical measurement setups and contributed the VNA and LDV data. H. K. H. peformed the FIB sample preparations, TEM, SAED and EDX data collections and analyses. M. X. and S. F. D. S. contributed the XRD data. T. L. and W. C. assisted on VNA data analyses. The rest of the authors contributed valuable opinions through active discussions.

**Data Availability Statement**

Data is available for sharing upon request with the corresponding authors.

**Supporting Information**

Supporting Information is available from the Wiley Online Library or from the author.

**Cladding Layer Enhanced GHz Bulk Acoustic Wave Resonance in Sodium Niobate Thin Films on Silicon**

We fabricate $NaNbO_3$-based solidly mounted bulk acoustic resonators on silicon (001) wafer without Bragg reflector and acoustic cavity. With the assistance of $LaAlO_3$ cladding layers, the vertical ferroelectric anisotropy of the stack is enhanced, resulting in a strong resonance signal and effective piezoelectric coefficient.

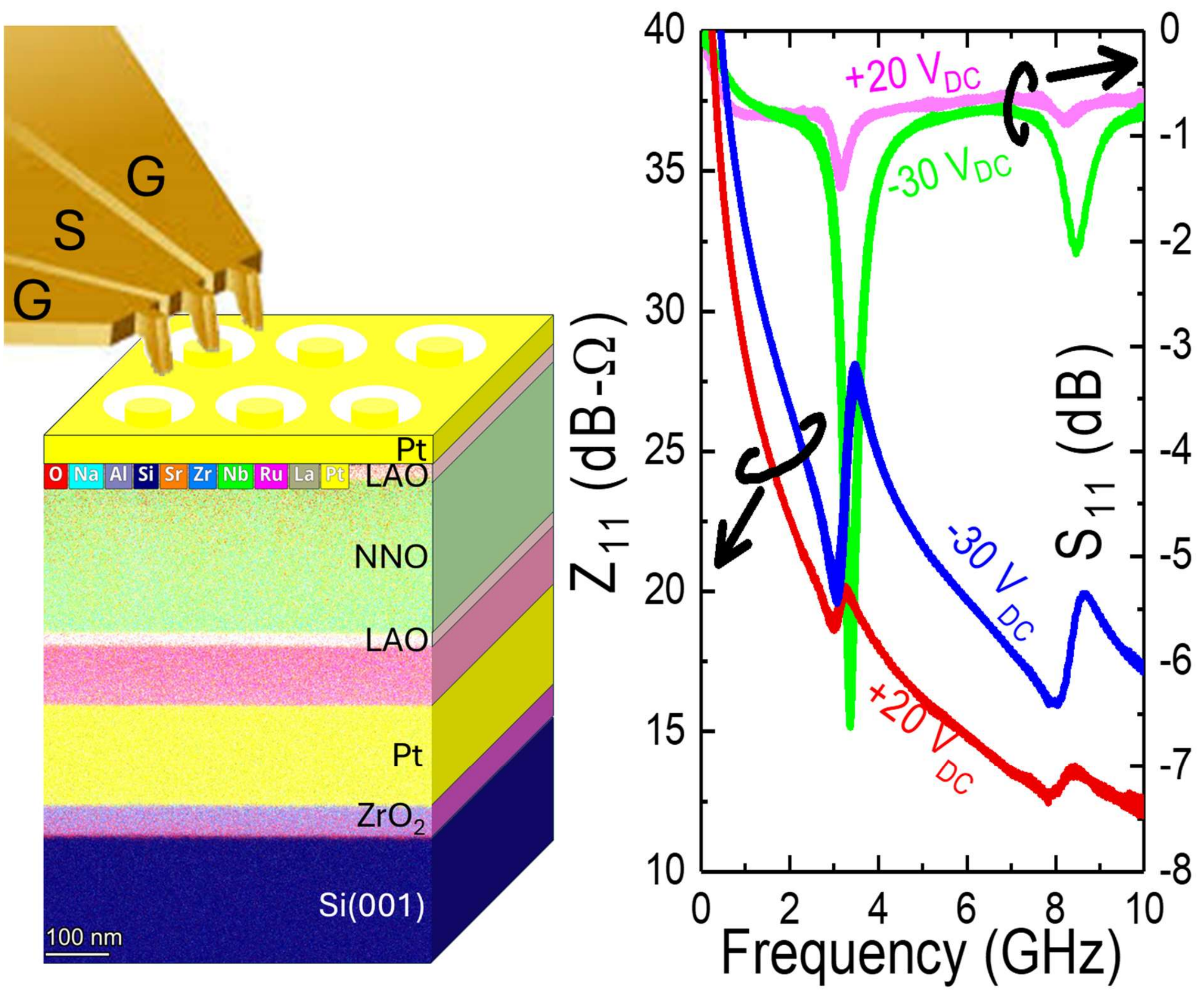

Dimension chosen: 55 mm broad × 50 mm high (1467 pixels × 1333 pixels)

Aspect ratio: 1.1

# Supporting Information

## Cladding Layer Enhanced GHz Bulk Acoustic Wave Resonance in Sodium Niobate Thin Films on Silicon

*Zhi Shiuh Lim*, Qibin Zeng, Hui Kim Hui, Mengyao Xiao, Tiancheng Luo, Weifan Cai, Shengwei Zeng, Samantha Faye Duran Solco, Baichen Lin, Celine Sim, Zhen Ye, Jinlong Xu, Mingxi Chen, Wei Fu, Chee Kiang Ivan Tan, Seeram Ramakrishna, Yeng Ming Lam, Vincent Chengkuo Lee, Ariando Ariando, Huajun Liu**

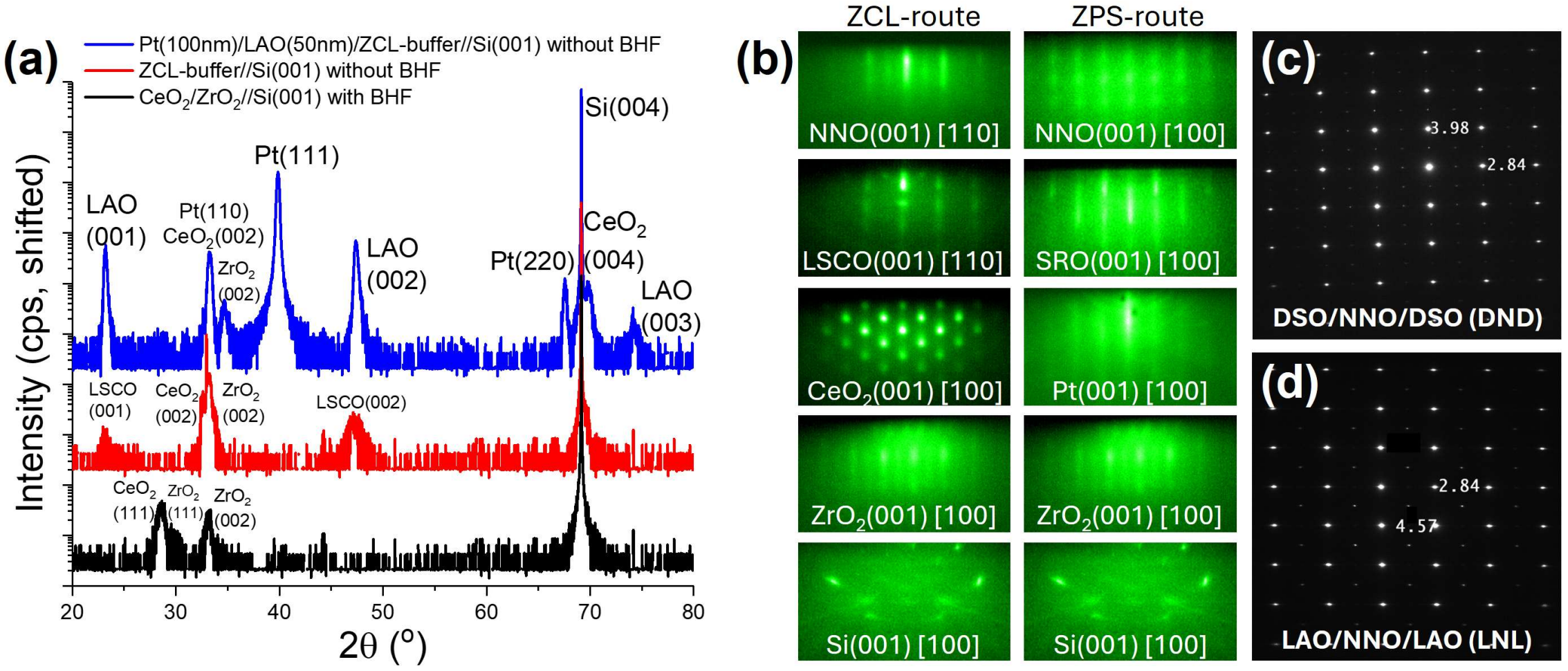

**Supporting Figure S1: (a)** Comparison between usage with (black curve) and without (red curve) BHF in growing the ZCL-buffer. The usage of BHF resulted in appearance of undesirable $ZrO_2$(111) $CeO_2$(111) phases but amorphous LSCO. Blue curve: growing Pt at elevated temperature on LAO(002)/ZCL-buffer//Si(001) resulted in an undesirable Pt(111) phase but absence of Pt(002) phase. **(b)** RHEED pattern evolutions of the ZCL- and ZPS routes with the substrate corner (Si[100]) parallel to the electron beam direction. **(c)-(d)** Measurement of lattice parameters of the DND and LNL heterostructures by the TEM-SAED within an area of 100 nm diameter (probe beam size) of the NNO layer's cross section, with the zone axis along Si[110]. The c-axis lattice parameters of NNO were found to be 3.98 Å and 4.57 Å for DND and LNL respectively, while the a-axis lattice parameter is $2.84*\sqrt{2} = 4.016$ Å for both.

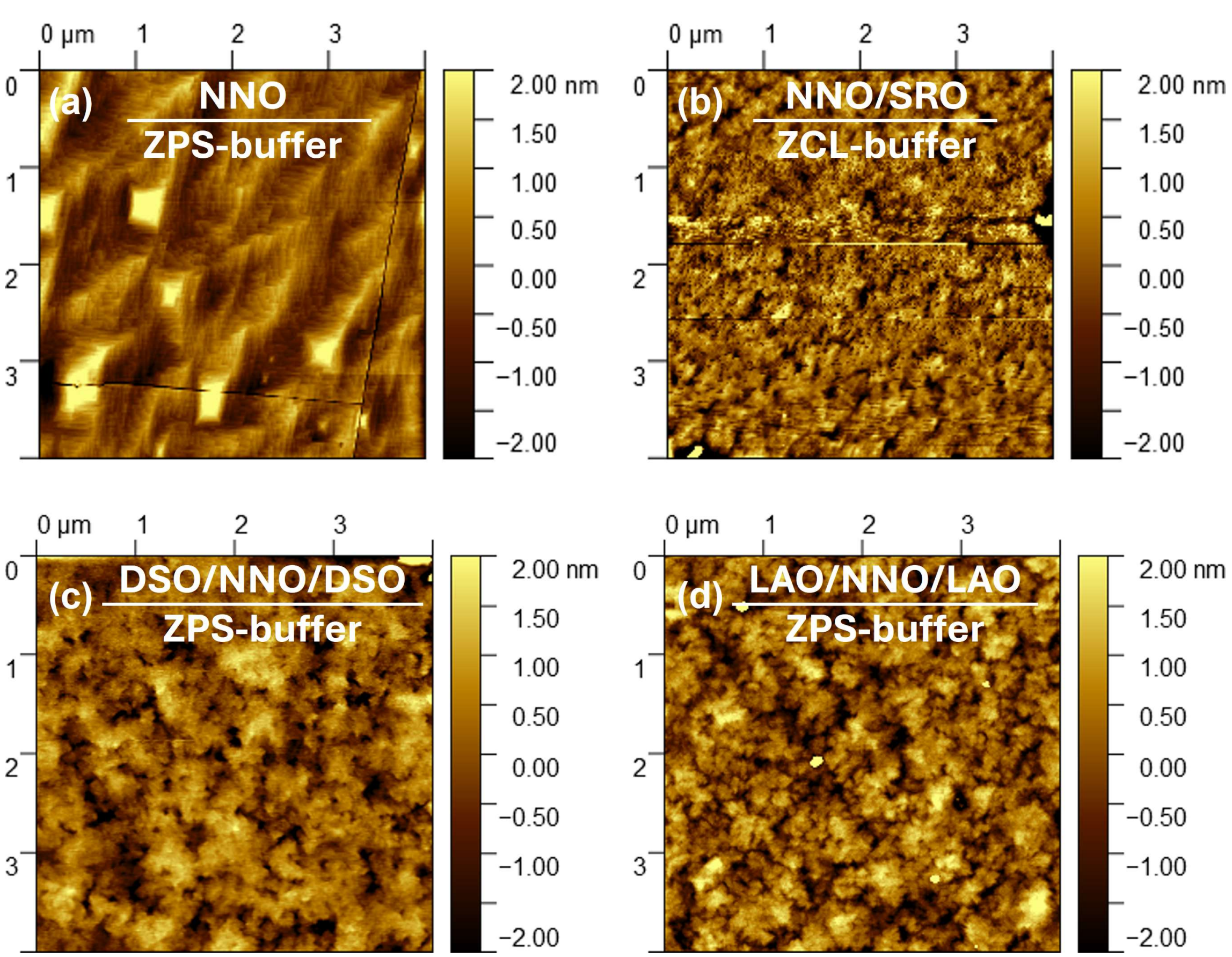

**Supporting Figure S2: (a)-(d)** Topography scans within 4-μm areas corresponding to the heterostructures listed in main text figures 3b-e before the photolithography, top Pt deposition and lift-off steps. Cracks can be seen in **(a)**, consistent to main text figure 3b but not in **(b)-(d)**. The r.m.s. roughness values are **(a)** 0.42 nm, **(b)** 0.78 nm, **(c)** 0.73 nm and **(d)** 1.2nm respectively.

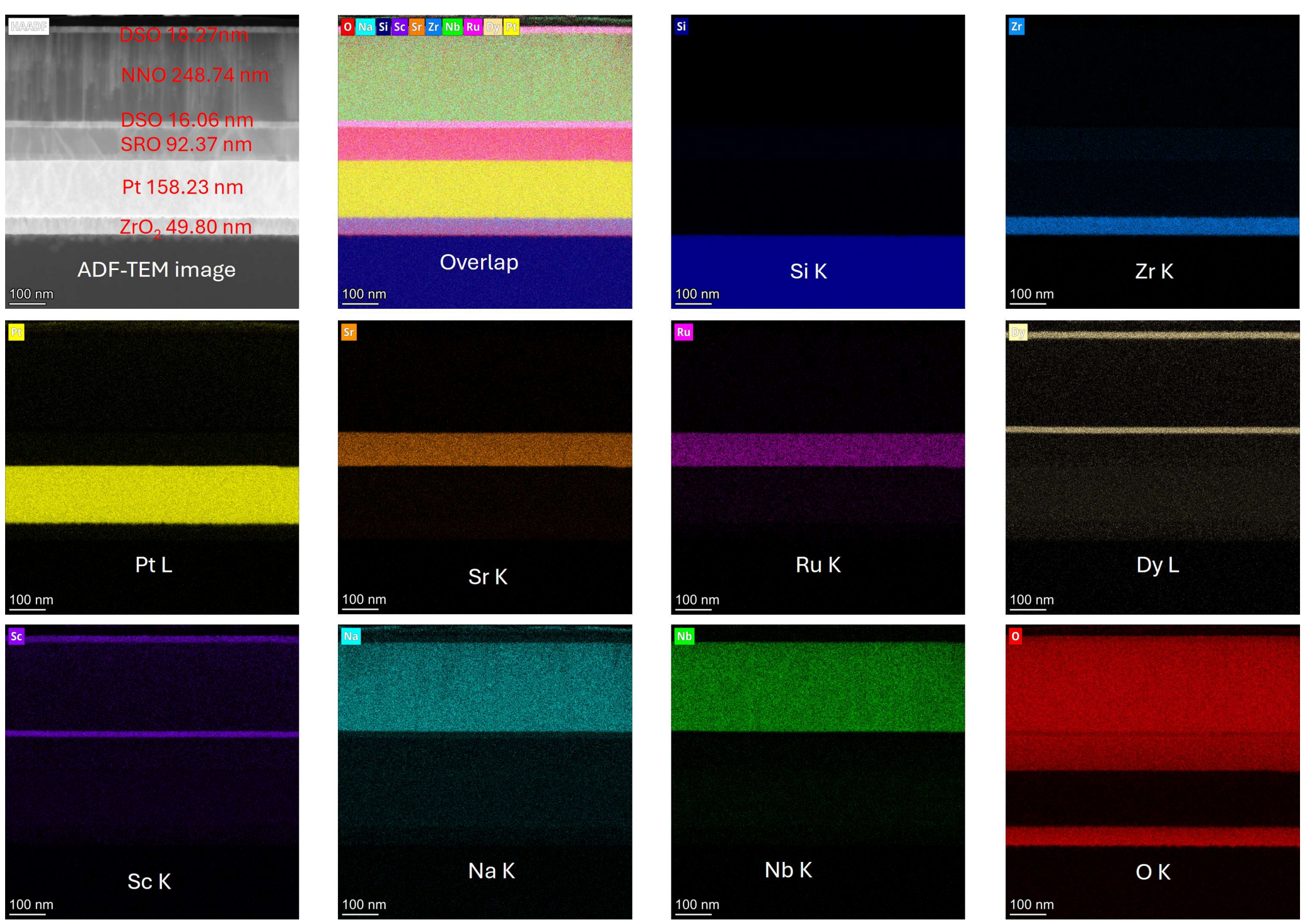

**Supporting Figure S3:** Cross-sectional element-specific mappings for the DSO/NNO/DSO (DND) heterostructure done by TEM-EDX. The annular dark-field (ADF) imaging with thickness measurements are shown at the top left corner. The layers are mapped by the core level X-ray emission from the Si, Zr, Pt, Sr, Ru, Dy, Sc, Na, Nb and O elements respectively.

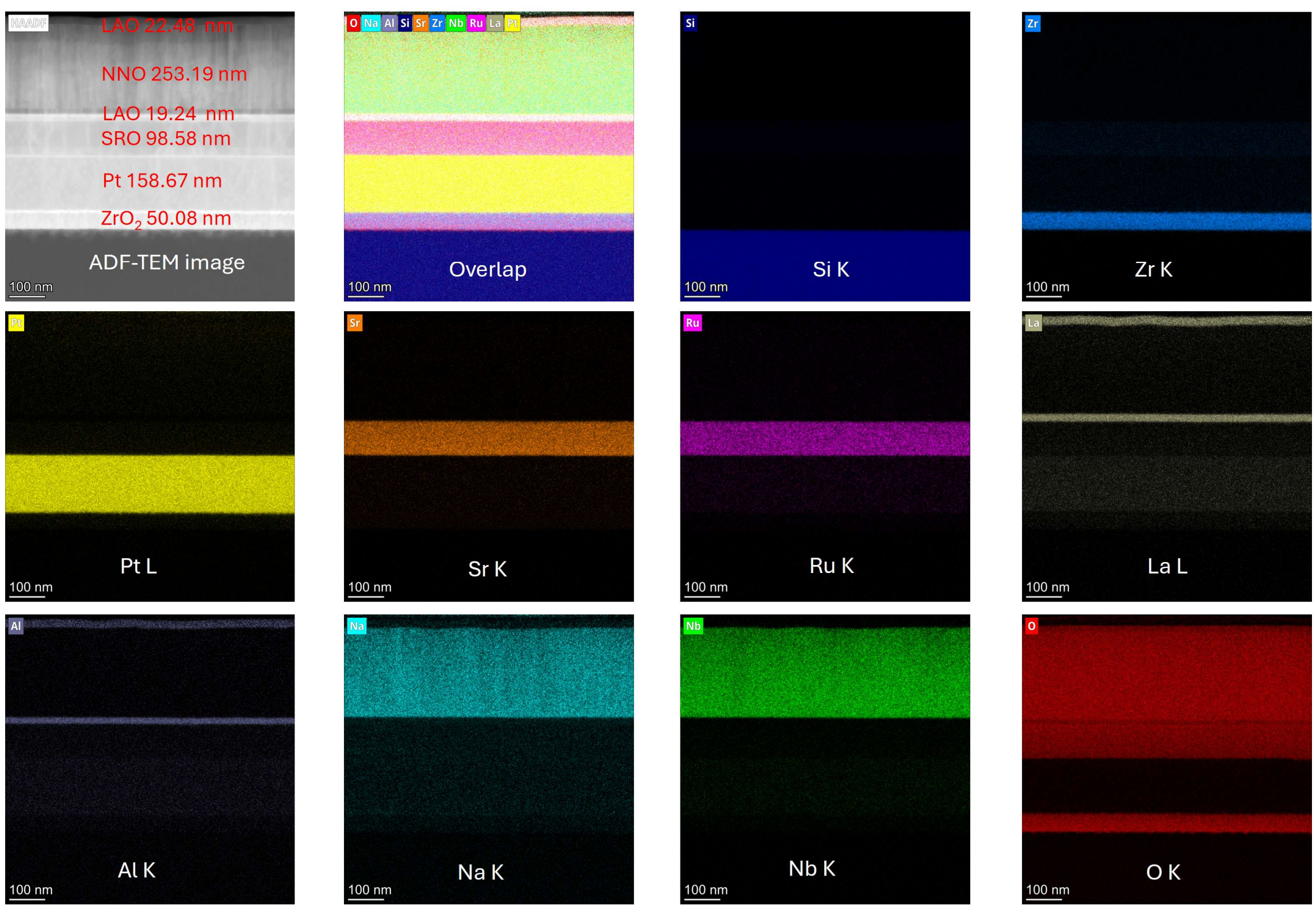

**Supporting Figure S4:** Cross-sectional element-specific mappings for the LAO/NNO/LAO (LNL) heterostructure done by TEM-EDX. The annular dark-field (ADF) imaging with thickness measurements are shown at the top left corner. The layers are mapped by the core level X-ray emission from the Si, Zr, Pt, Sr, Ru, La, Al, Na, Nb and O elements respectively.